\documentstyle[12pt]{article}
\newcommand  {\arr} [2]{\begin{array}{#1}#2\end{array}}

\renewcommand{\bar} [1]{\overline{#1}}
\newcommand  {\clos}[2]
    {\mathop{\mbox{clos}}_{{\scriptstyle #1}}{\textstyle\left\{#2\right\}}}
\newcommand  {\cnjg}[1]{\overline{#1}}

\renewcommand{\d}      {{\rm d}}

\newcommand  {\dd}  [2]{\frac{\partial #2}{\partial #1}}
\newcommand  {\ds}     {\displaystyle}
\newcommand  {\ea}  [1]{\begin{eqnarray}#1\end{eqnarray}}
\newcommand  {\ep}     {\epsilon}
\newcommand  {\eq}  [1]{\begin{equation}#1\end{equation}}
\newcommand  {\F}      {{\cal F}}
\newcommand  {\Fu}  [2]{\hat{u}_{#2}(\v{#1})}
\newcommand  {\h}   [2]
       {\mbox{\boldmath $h$}_{#2}(\mbox{\boldmath $#1$})}
\renewcommand{\hat} [1]{\widehat{#1}}
\renewcommand{\i}      {{\rm i}}
\renewcommand{\P}      {{\cal P}}
\newcommand  {\ph}     {\varphi}
\newcommand  {\R}   [1]{{\bf R}^{#1}}
\newcommand  {\Rz}     {{\bf R}^{3}\backslash\{\v 0\}}
\newcommand  {\Sum} [1]{\sum_{{\scriptstyle #1}}}
\newcommand  {\T}   [1]{{\bf T}^{#1}}
\newcommand  {\th}     {\vartheta}
\newcommand  {\thatis} {\Longleftrightarrow}
\newcommand  {\then}   {\Longrightarrow}
\renewcommand{\tilde}  {\widetilde}
\renewcommand{\v}   [1]{\mbox{\boldmath $#1$}}

\newcommand  {\vFu} [1]{\v{\hat{u}}(\v{#1})}
\newcommand  {\vr}  [1]{\v e_r(\v{#1})}
\renewcommand{\vss} [1]{\mbox{{\scriptsize\boldmath$#1$}}}
\newcommand  {\vth} [1]{\v e_\vartheta(\v{#1})}
\newcommand  {\vph} [1]{\v e_\varphi(\v{#1})}
\newcommand  {\Z}   [1]{{\bf Z}^{#1}}
\newcommand  {\Zz}     {{\bf Z}^{3}\backslash\{\v 0\}}

\newtheorem  {theorem}{Theorem}
\newcommand  {\Theorem}[2]
              {\begin{theorem}[#1]{\sl #2}\end{theorem}}
\newtheorem  {corollary}[theorem]{Corollary}
\newcommand  {\Corollary}[2]
           {\begin{corollary}[#1]{\sl #2}\end{corollary}}
\newtheorem  {proposition}[theorem]{Proposition}
\newcommand  {\Proposition}[2]
           {\begin{proposition}[#1]{\sl #2}\end{proposition}}
\newtheorem  {postulation}[theorem]{Postulation}
\newcommand  {\Postulation}[2]
           {\begin{postulation}[#1]{\sl #2}\end{postulation}}
\newcommand  {\Proof}[1]{{\sc Proof}: #1$\Box$\vspace{6pt}}

\newcommand{\sakujo}[1]{}

\begin{document}

\begin{center}
{\Large \bf 
Multiresolution approximation 
of the vector fields on $\T3$
}\vspace{12pt}\\
\newcommand{\AUTHOR}[2]{{\bf #1}\\{#2}\vspace{6pt}\\}
\AUTHOR
  {ARAKI Keisuke}
  {Department of Mechanical Engineering, Okayama University 
   of Science,\\ Okayama 700-0005, JAPAN}
\AUTHOR
  {SUZUKI Katsuhiro}
  {Department of Applied Physics, Tokyo University of 
   Agriculture and Technology,\\ Fuchu 183-0054, JAPAN}
\AUTHOR
  {KISHIDA Keiji}
  {Department of Material Science, Hiroshima University,\\ 
   Higashi-hiroshima 739-8526, JAPAN}
\AUTHOR
  {KISHIBA Seigo}
  {Information Processing Center, Hiroshima University,\\ 
   Higashi-hiroshima 739-8526, JAPAN}
\end{center}

\section*{abstract}

Multiresolution approximation (MRA) of the vector fields on 
$\T3$ is studied.
We introduced in the Fourier space a triad of vector fields 
called helical vectors which derived from the spherical 
coordinate system basis.
Utilizing the helical vectors, we proved the orthogonal 
decomposition of $L^2(\T3)$ which is a synthesis of the Hodge 
decomposition of the differential 1- or 2-form on $\T3$ and 
the Beltrami decomposition that decompose the space of 
solenoidal vector fields into the eigenspaces of curl 
operator.
In the course of proof, a general construction procedure 
of the divergence-free orthonormal complete basis from the basis 
of scalar function space is presented.
Applying this procedure to MRA of $L^2(\T3)$, we discussed 
the MRA of vector fields on $\T3$ and the analyticity and 
regularity of vector wavelets.
It is conjectured that the solenoidal wavelet basis must 
break $r$-regular condition, i.e. some wavelet functions cannot be 
rapidly decreasing function because of the inevitable singularities of 
helical vectors.
The localization property and spatial structure of  
solenoidal wavelets derived from the Littlewood-Paley type 
MRA (Meyer's wavelet) are also investigated numerically.

\newpage
\baselineskip 24pt
\section{Introduction}

Divergence free vector fields with coherent structures are 
ubiquitous in a lot of natural phenomena, for example, coronal 
flare of the Sun, dipolar magnetic field of the Earth, or the 
coherent vortices in sufficiently subsonic motions of fluid,
for example Great Red Spot in Jupiter.
Wavelet analysis has been regarded one of the promising tools 
for surveying such coherent structures.
Because they are remarkably localized not only in physical 
space but also in Fourier space within the limit of the 
uncertainty principle.
Using wavelet transformation, one can obtain the information 
of scale and location simultaneously.

Theory of discrete wavelet transformation is well known as 
multiresolution approximation (MRA) of function spaces.
The wavelet bases are remarkably useful because 
they are the orthonormal complete, for some cases 
unconditional, basis not only of $L^2$ space but also of many 
function spaces such as Sobolev spaces, H\"older spaces, Hardy 
space, Besov spaces, {\it etc} \cite{meyer1992}.
Dyadic dilation property of  the wavelet basis seems quite 
akin to the idea of scaling laws, which appears in many 
fields of physics such as quantum field theory \cite{feder}, 
critical phenomena \cite{ma} or fully developed turbulence 
\cite{frisch}.

The wavelets, however, are scalar function so that application 
to divergence-free vector field contains a problem.
Surely one can apply the scalar-valued wavelet transform to 
each component of a divergence-free vector field, say $\v u(
\v x)$ = ( $u_x(x,y,z)$, $u_y(x,y,z)$, $u_z(x,y,z)$ ),
 and obtain the wavelet spectrum of the field, 
\eq{
  \v u(\v x) =    \sum_\lambda\left( 
 \bigl<\psi_\lambda,u_x\bigr>\psi_\lambda(\v x), 
 \bigl<\psi_\lambda,u_y\bigr>\psi_\lambda(\v x), 
 \bigl<\psi_\lambda,u_z\bigr>\psi_\lambda(\v x)
   \right),
}
where $\psi_{\lambda}$ is scalar wavelet and 
$\left\langle *, * \right\rangle$ denotes inner product. 
Each term of the spectrum 
\eq{
  \v u_\lambda(\v x)=\left(
   \bigl<\psi_\lambda,u_x\bigr>\psi_\lambda(\v x), 
   \bigl<\psi_\lambda,u_y\bigr>\psi_\lambda(\v x), 
   \bigl<\psi_\lambda,u_z\bigr>\psi_\lambda(\v x)
  \right)
}
is not a divergence-free vector field in general.
This discrepancy has its root in the fact that only two of 
three components are independent, but on the whole they are 
dependent each other.
(Furthermore they depends also on boundary conditions.)
Thus the divergence-free vector-valued wavelet function is 
required for practical purposes.

Divergence-free vector wavelet bases have been proposed by Battle 
and Federbush\cite{battle1993}, and Frick and Zimin\cite
{frick1993}.
Battle and Federbush adopted such a constructing way that 
minimizes the value of integral $\int(\nabla\times\v A)^2\d \v x$
 under the two constraints, the divergence-free 
condition and an appropriate boundary condition.
On the other hand, Frick and Zimin proposed such a wavelet 
that is, roughly speaking, given by the curl of the function given by Fourier 
integral of a step function supported on a spherical shell 
in the Fourier space. 
The former approach requires the variational calculations 
when the wavelet transformation is carried out.
The latter approach, on the other hand, has no such steps 
because it is based on the sharp decomposition of Fourier 
space into spherical shells. 
The obtained wavelet functions, however, are not orthogonal 
each other in general, and not localized well, in other words, 
breaks $r$-regular condition.
Because of these discrepancies, the approaches they 
proposed do not seem popular in practical applications.

In the present work, we propose a general construction 
procedure of the orthonormal complete divergence-free vector 
wavelet basis of $L^2(\T3)$, in which only two popular 
algorithms, fast Fourier transform (FFT) and fast wavelet 
transform (FWT), and no additional novel one is required.

The procedure we will describe here is based on a quite 
different idea from the previous two.
    There are two key ideas for the construction.
    One is that any function expansion using an orthonormal 
complete basis, $\{f_\lambda;\lambda\in\Lambda\}$ (say), is a 
unitary transform from $L^2(\T3)$ to $l^2(\Lambda)$, where 
$\Lambda$ is an appropriate set of indices.
    Therefore the Fourier coefficients of the base functions 
$\hat f_\lambda(\v k)$ are regarded as components of an 
infinite dimensional unitary matrix which acts on $l^2$ space 
which maps Fourier coefficients to 
$\{f_\lambda\}$-expansion coefficients.
    The other one is that all the orthonormal complete basis 
of the function space of solenoidal vector fields on $\T3$,
which is denoted by $L^2_\Sigma(\T3)$ hereafter, is given by 
a certain unitary transform of the complex helical wave basis
\cite{moffatt,lesieur1997}.
    Thus functions to which complex helical waves are unitary 
transformed by the matrix $\bigl\{\F \{f_\lambda\}\bigr\}$ 
constitutes an another orthonormal basis of $L^2_\Sigma(\T3)$.
    As an orthonormal complete basis of $L^2(\T3)$, wavelet 
basis is adopted here.

This study is an attempt to construct 
the multiresolution approximation of the vector fields.
In the present study we restrict our interest on the vector 
fields on the three-torus $\T3$ and the possibility of 
construction of multiresolution approximation of them.
The reason of choice of the manifold $\T3$ is mainly due to 
the fact that the Hodge decomposition theorem is established 
on bounded manifolds.
Thus we base our attempt on the MRA of $L^2(\T3)$ though 
the theory of wavelets is firstly established on the 
unbounded Euclidean space $\R{N}$.

This paper is organized as follows.
In \S2 notations in the paper are explained.
The orthogonal decomposition of the vector field on $\T3$ is 
proved in \S3.
Algorithm of helical wavelet decomposition is given in \S4, 
which is an anthology of the properties of helical basis.
Construction the theory of MRA of vector fields is tried 
in \S5.
The Riesz basis condition, which is one of the basic 
properties of MRA, is shown to be broken.
Section 6 is devoted to the discussion on the {\it regularity}, 
ie the localization property of the helical wavelet.
Finally some remarks are given in \S7.

\section{Nomenclature}

Before going into the details, some notations should be fixed.

 Let us denote by ${\bf R}^3$ a  linear 
space spanned by a  Cartesian basis $\{ \v e_x, \v e_y, \v e_z \}$,  
and  $\T3$ the quotient space  ${\bf R}^3 / \Z3$. 
$\v\chi(\T3)$ is a set of   vector fields on $\T3$ defined as 
 $${\v \chi(\T3)}
:= \biggl\{ {
{\v u}\  ; \ \ 
{\v u}({\v x})=\sum\limits_{i=x,y,z} {u_i({\v x})} {\v e}_i, \ 
\  u_x,u_y,u_z \in C^{\infty}(\T3)\  
}
\biggr\}, 
$$ 
 where 
 $\{{\v e}_x, {\v e}_y,  {\v e}_z \}$ is the basis of 
the tangent space $T_{\v x} \T3$ at the point  ${\v x} \in \T3$ 
obtained by the canonical identification. 
%
In other words,  $\v\chi(\T3)$ is the set of all 
the $C^{\infty}$ sections of the  tangent  bundle $T\T3$.
 We identify, if required,
 the space with the set of 1-forms   $\Omega^1(\T3)$,
 or that of 2-forms $\Omega^2(\T3)$.

%
%

The completion of $F$ with respect to the norm of 
the Banach space $E$ is denoted by $\clos{E}{F}$.
In the following analysis,  
we study the function space given by the $L^2$-norm completion of ${\v \chi}(\T3)$, 
\ea{L^2_{\v \chi}(\T3)
& := & \clos{L^2(\T3)}{\v\chi(\T3)},
\\ & = & 
\biggl\{ {
{\v u}\  ; \ \ 
{\v u}({\v x})=\sum\limits_{i=x,y,z} {u_i({\v x})} {\v e}_i, \ 
\  u_x,u_y,u_z \in L^{2}(\T3)\  }
\biggr\},  
}
which  is a Hilbert space equipped with a {\it inner product},  
\eq{\label{eq:inner product}
   \bigl< \v u, \v v \bigr>_{\vss\chi} 
   := \int_{\T3}
      \bar{\v u}(\v x)\cdot\v v(\v x)
   \,\d\v x
   = \int_{\T3} \Bigl(
     \bar{u_x(\v x)}v_x(\v x) + \bar{u_y(\v x)}v_y(\v x) +
     \bar{u_z(\v x)}v_z(\v x)
   \Bigr)\,\d\v x,
}
where $\v u$, $\v v \in L^2_{\vss{\chi}}(\T3)$, 
 $\cdot$ denotes the scalar product of two vectors, 
 $\bar *$  the complex conjugate,  and 
 $\d\v x$  the Lebesgue measure on $\T3$.
By definition,  any Hilbert space is a Banach space 
 with the  norm naturally determined by its inner product. 
Concerning $L^2_{\v \chi}(\T3)$, the norm is 
\eq{
 ||\v u|L^2_{\vss{\chi}}(\T3)|| 
:=
   \sqrt{ \bigl< \v u, \v u \bigr>_{\vss \chi} } 
=
   \sqrt{ \int_{\T3}\Bigl(
     |u_x(x,y,z)|^2 + |u_y(x,y,z)|^2 + |u_z(x,y,z)|^2
   \Bigr)\,\d\v x },
}
where $||*|E||$ denotes the norm of a Banach space $E$. 
In the following, 
we drop the symbol $\T3$ in definite integrals, 
and represent $\T3$ by a  periodic unit cube, i.e. 
 $\T3=[0,1]^3$. 

 Fourier series representation of $f({\v x}) \in L^2_{\vss \chi}(\T3)$ 
 is formally written as 
\eq{
   f(\v x)
      = \sum_{\vss k\in\Z3} 
        \hat f(\v k) \exp(2\pi\i\v k\!\cdot\!\v x),
}
where $\hat{f}({\v k})$'s  (${\v k}\in \Z3$) are 
 Fourier coefficients. 
The calligraphic letter $\F$ is used to denote 
the sequence of Fourier coefficients, i.e.  
 $\F f= \bigl \{ \hat{f}(\v k); \v k \in \Z3 \bigr\}$.
 It is also used for the Fourier 
 transform of a set of function 
 and a function space, for example
 $ \F L^2(\T3) = \bigl\{ \F f\, ;\, f \in L^2(\T3) \bigr\}.$

Let us consider a trivial bundle  ${\v M}:=\Z3 \times {\v R}^3$.  
{\it Fourier transform of a vector field} is defined by the 
Fourier transform of the components with respect to the 
Cartesian basis 
 as follows: 
\eq{
   \hat{\v{u}}(\v k) := 
   \hat u_x(\v k){\hat{\v e}}_x + \hat u_y(\v k){\hat{\v e}}_y + 
   \hat u_z(\v k){\hat{\v e}}_z,
}
 where  ${\v k}\in \Z3$ and 
$\left\{ {\hat {\v e}}_x, {\hat {\v e}}_y, {\hat {\v e}}_z \right\}$ 
 is  a Cartesian basis   of  the fiber ${\v R}^3$ of $\v M$. 
Thus $({\v k},{\v u}({\v k}))\in {\v M}$ and 
$\F\v u$ is a sequence of three dimensional vectors. 

Being introduced the $l^2_{\vss{\chi}}$-norm of $\F L^2_{\vss
\chi}(\T3)$ 
 defined by 
\eq{
   ||\F\v u|l^2_{\vss{\chi}}||:=\sqrt{
   \sum_{\vss k\in\Z3}\Bigl(
      |\hat{u}_x(\v k)|^2 + |\hat{u}_y(\v k)|^2 + 
      |\hat{u}_z(\v k)|^2
   \Bigr)},
}
 $\F L^2_{\vss \chi}(\T3)$
becomes a Banach space.
Applying Parseval identity to each Cartesian component 
$u_x$, $u_y$ and $u_z$ of a $L^2_{\vss\chi}(\T3)$ vector 
field $\v u$, we conclude that the $L^2_{\vss\chi}(\T3)$ 
norm and $l^2_{\vss{\chi}}(\Z3)$ norm are equivalent.
Due to this equivalence of two norms, the {\it fundamental 
sequence} of $\v u(\v x)$ defined by
\eq{
   \v u_N(\v x):=
      \sum_{0\leq|\vss k|<N}
      \hat{\v{u}}(\v k) \exp(2\pi\i\v k\cdot\v x),
   \ N\in{\bf N}.
}
is a Cauchy sequence of vector fields in the sense of 
$L^2_{\vss\chi}$-norm.
   In the following, the derivatives 
 are formally defined by Fourier series,   
\eq{
   \dd{x_j}{u_i} := \frac{1}{(2\pi)^3}
      \Sum{\vss k\in\vss\Zz}
      2\pi\i\,k_j\,\hat{u_i}(\v k)
      \exp(2\pi\i\v k\cdot\v x),
}
where $i$, $j=x$, $y$ and $z$.

    We distinguish the terms {\it divergence-free} and 
{\it solenoidal} in the present work;
    the former is used for such vector fields $\v u$ that satisfy 
$\nabla\cdot\v u=0$.
    The latter term, on the other hand, is used when a 
vector field $\v u$ is given by curl of certain vector field $\v a$: 
$\v u = \nabla\times\v a$. 
 Difference of these two kinds of vectors persists in the 
fact that the harmonic vector fields, which are constant function 
for the case of $\T3$, are also divergence-free. 
In terms of forms, 
 divergence-free fields  correspond to  {\it closed 2-forms}  
and solenoidal fields {\it exact 2-forms}. 

%
\section
    {orthogonal decomposition of the vector fields on $\T3$}

In order to construct complex helical waves, a triad of 
vector fields $\bigl\{ \vr{\v k}, \vth{\v k}, \vph{\v k} 
\bigr\}$ which is almost identical to a spherical coordinate 
system basis is introduced in the Fourier space.
In the present study, they are defined in terms of the 
wavenumber vector $\v k$ and Cartesian coordinate system 
basis \{$\v e_x$,$\v e_y$,$\v e_z$\} as follows:
\eq{
   \vr{k}:=\frac{\v k}{|\v k|},\ \ \ 
   \vph{k}:=\left\{ \arr{cc}{\ds
      \vr{k}\times\v e_x & (\vr{k}/\!\!/\v e_z)
      \vspace{9pt}\\\ds
      \frac{\v e_z\times\vr{k}}
           {|\v e_z\times\vr{k}|}
      & (\mbox{otherwise})
   },\right.\ \ \ 
   \vth{k}:=\vph{k}\times\vr{k}.
}
The helical vectors $\h{k}{s}$, where the index $s$ denotes 
{\it polarity} of them and is $+$, $-$ or $0$, are a triad 
of complex valued vector fields on the Fourier space that 
are defined by 
\eq{\label{eqn:helical vectors}
   \h{k}{+}:=\frac{\vth{k}+{\rm i}\vph{k}}{\sqrt{2}},\ \ 
   \h{k}{-}:=\frac{\vth{k}-{\rm i}\vph{k}}{\sqrt{2}},\ \ 
   \h{k}{0}:=-\i\vr{k}.
}
We also use the notations $\h{k}{\Sigma+}$, $\h{k}{\Sigma-}$ 
and $\h{k}{D}$ instead of $\h{k}{+}$, $\h{k}{-}$ and $\h{k}
{0}$, respectively.

It should be noted here that we used definition of helical 
vectors which is used in Ref.\cite{waleffe1992} 
with slight modifications, introduction of $\h{k}{0}$ and 
normalization of $\h{k}{\pm}$ vectors.
The helical vectors are defined on $\Rz$, and infinitely 
differentiable vector fields on $\Rz$ except on the line 
along the north pole and the south pole for $x$ and $y$ 
component of $\h{k}{\pm}$ and around $\v k=\v 0$ for $z$ 
components of $\h{k}{\pm}$ and all the components of $\h{k}
{0}$.
In this section, however, we treat the vector field on $\T3$ 
and restrict the case for $\v k\in\Zz$.

Utilizing the helical vectors, we define the {\it complex 
helical waves} that are complex valued vector fields by
\eq{
   \v V(\v k,s;\v x):= \h{k}{s} \exp(2\pi\i\v k\cdot\v x),
}
for $\v k\in\Zz$ and $s=+$, $-$, $0$.
According to the polarity of the helical vectors, we will 
call each helical wave $\Sigma+$-, $\Sigma-$- and $D$-mode, 
respectively.
It should be remarked that we introduced $D$-mode of 
complex helical waves that is not defined by Lesieur\cite
{lesieur1997} or Waleffe\cite{waleffe1992}.
We will see in the following that this mode is curl-free.

The helical vectors are defined to satisfy the orthonormal 
relation 
\eq{
   \cnjg{\h{k}{s}}\cdot\h{k}{s'}=\delta(s|s'),
}
for $s$, $s'=+$, $-$, $0$ at each $\v k$.
Hereafter $\delta(A|B)$ denotes Kronecker's delta whose value 
is one only when the two arguments $A$ and $B$, which are not 
only numbers but also vectors, symbols, etc., coincide, 
otherwise it gives zero.
Therefore the complex helical waves are such vector fields 
that are orthonormal in $L^2_{\vss\chi}(\T3)$:
\eq{
   \bigl< \v V(\v k,s;\v x), \v V(\v k',s';\v x) \bigr>_{\vss\chi}
   = \delta\left(\v k,s\left|\v k',s'\right.\right).
}
Scalar and vector products of the wavenumber vector and the 
helical vectors are 
\ea{&&
   \i\v k\cdot\h{k}{\pm}=0,
\ \ \ 
   \i\v k\cdot\h{k}{0}=|\v k|.
\\&&
   \i\v k\times\h{k}{\pm}= \pm|\v k|\h{k}{\pm},
\ \ \ 
   \i\v k\times\h{k}{0}= \v 0,
}
    Corresponding to these products, divergence and curl of 
complex helical waves are 
\ea{&&
   \nabla\cdot\v V(\v k,\pm;\v x)=0,
\ \ \ 
   \nabla\cdot\v V(\v k,0;\v x)=
      2\pi|\v k|\exp(2\pi\i\v k\cdot\v x),
\\&&
   \nabla\times\v V(\v k,\pm;\v x)=
      \pm 2 \pi|\v k|\v V(\v k,\pm;\v x),
\ \ \ 
   \nabla\times\v V(\v k,0;\v x)=\v0.
}
    The $\Sigma\pm$-modes are eigenfunctions of the curl 
operator which are associated with the eigenvalues $ \pm 2 \pi 
|\v k| $ and the $D$-mode belongs to the kernel of the 
curl operator.
    Thus it is obvious now that the indices of helical vectors 
$+$, $-$ and $0$ correspond to the signs of eigenvalues, and 
that the implication of the alternative indices $\Sigma+$, 
$\Sigma-$ and $D$ are {\it solenoidal with positive 
helicity}, {\it solenoidal with negative helicity} and 
{\it dilatational}, respectively.

    Since $\vr{-k}=-\vr{k}$, $\vth{-k}=\vth{k}$, $\vph{-k}=
-\vph{k}$, the complex conjugates of helical vectors at $\v k$ 
are equal to ones at $-\v k$:
\eq{
   \bar{\h{k}{s}}=\h{-k}{s}. 
}
Therefore, complex conjugates of the complex helical waves 
satisfy 
\eq{
   \cnjg{\v V(\v k,s;\v x)}=\v V(-\v k,s;\v x),
}
for $\v k\in\Zz$, $s=+$, $-$, $0$.
The $\Sigma+$ and $\Sigma-$ vectors are complex conjugate 
each other,
\eq{
   \bar{\h{k}{\pm}}=\h{k}{\mp},
}
so that the $\Sigma\pm$-modes of the complex helical waves 
are complex conjugate each other:
\eq{
   \cnjg{\v V(\v k,\pm;\v x)}=\v V(\v k,\mp;\v x).
}

Using these helical vectors, we define the {\it helical 
decomposition of Fourier coefficients (or $l^p_{\vss\chi}
$-sequence)}, say $\{\hat{\v{u}}(\v k)\}$, by 
\eq{
   \hat{\v{u}}(\v k) = 
   \hat{u}_+(\v k) \h{k}{+} + 
   \hat{u}_-(\v k) \h{k}{-} + 
   \hat{u}_0(\v k) \h{k}{0},
}
where $\hat u_s(\v k)$'s are the {\it s-mode helical Fourier 
coefficients} defined by the scalar product 
\eq{
   \hat u_s(\v k) := \hat{\v u}(\v k)\cdot\cnjg{\h{k}{s}}, 
   \label{eq:helical fourier coefficient}
}
for each $\v k\in\Zz$ and $s=+$, $-$, $0$.
    Each helical Fourier coefficients $\{\hat u_s(\v k)\}$ 
is a sequence of scalars.

Conversely, by multiplying helical vector on each terms of a 
scalar ($l^p$-)sequence $\{\hat{u}(\v k);\v k\in\Zz\}$, one 
can make a vector sequence $\{\hat{u}(\v k)\h{k}{s};\v k\in
\Zz\}$ which we call the {\it helical pull up of a sequence 
$\{\hat{u}(\v k)\}$ to $s$-mode } hereafter.

    It should be remarked here that the helical decomposition 
of solenoidal vector field on $\T3$ has also been discussed 
by Constantin and Majda\cite{consta1988}.
    In their paper, the decomposition is given by $\ds 
\hat{\v{u}}_\pm(\v k) = \hat{\v{u}}(\v k) \pm \i \vr{k} 
\times \hat{\v{u}}(\v k)$ for each $\v k\in\Zz$ 
under the assumption that the vector field is solenoidal 
$\v k\cdot\hat{\v{u}}(\v k)=0.$
    It is easy to see that, under the same assumption, the 
identity $\ds \hat{\v{u}}_\pm(\v k) = \sqrt2\hat u_\pm(\v k)
\h{k}{\pm}$ is satisfied for each $\v k\in\Zz$.

The helical decomposition of Fourier coefficients is 
unitary in the sense that the identities 
\eq{
   \cnjg{\hat{\v{u}}(\v k)}\cdot\hat{\v{v}}(\v k) = 
   \cnjg{\hat{u}_+(\v k)} \hat{v}_+(\v k) + 
   \cnjg{\hat{u}_-(\v k)} \hat{v}_-(\v k) + 
   \cnjg{\hat{u}_0(\v k)} \hat{v}_0(\v k)
}
are satisfied for each $\v k\in\Zz$ if 
$\left\{\hat{\v{u}}(\v k)\right\}$ and 
$\left\{\hat{\v{v}}(\v k)\right\}$
belongs to $l^\infty_{\vss\chi}(\Zz)$.
This property leads to the identities
\eq{\label{eq:unitary seq.}
   |\hat{u}_x(\v k)|^2 + 
   |\hat{u}_y(\v k)|^2 + 
   |\hat{u}_z(\v k)|^2 = 
   |\hat{u}_+(\v k)|^2 + 
   |\hat{u}_-(\v k)|^2 + 
   |\hat{u}_0(\v k)|^2
}
for every $\v k\in\Zz$,
and to the following consequences.

\Proposition{convergence of $l^2$-sequences}
{
If a three-vector sequence $\{ \vFu{k} \}$ belongs to 
$l^2_{\vss\chi}(\Z3)$, 
each of its helical projections $\{ \Fu{k}{s} \}$ is 
$l^2(\Zz)$-sequence for $s=+$, $-$ and $0$.
Conversely, if an sequence $\{ \hat u(\v k) \}$ belongs to 
$l^2(\Zz)$, each of its helical pull up $\{ \hat u(\v k) 
\h{k}{s} \}$ is $l^2_{\vss\chi}(\Zz)$-sequence for $s=+$, 
$-$ and $0$.
}
\Proof{
equation (\ref{eq:unitary seq.}) leads to the inequalities
$$
   |\hat{u}_s(\v k)|^2 \leq
   |\hat{u}_x(\v k)|^2 + 
   |\hat{u}_y(\v k)|^2 + 
   |\hat{u}_z(\v k)|^2,
$$
for every $\v k\in\Zz$ and $s=+$, $-$ and $0$.
Thus the $l^2$-norm of $\{ \Fu{k}{s} \}$ satisfies
$$
   \Sum{\v k\in\Zz} \hspace{-9pt} |\hat{u}_s(\v k)|^2 
   \leq \Sum{\v k\in\Zz} \hspace{-9pt} \Bigl(
     |\hat{u}_x(\v k)|^2 + 
     |\hat{u}_y(\v k)|^2 + 
     |\hat{u}_z(\v k)|^2 
   \Bigr) 
   \leq ||\hat{\v{u}}|l^2_{\vss{\chi}}(\Z3)||
   < \infty.
$$
The latter part is proven by two steps.
First, the absolute values of the helical vectors are one, 
i.e. $|\h{k}{s}|=1$, so that all the Cartesian components of 
them are equal to or less than one, i.e. $|(\h{k}{s})_j|\leq1$,
for every $\v k\in\Zz$, $s=+$, $-$, $0$, and $j=x$, $y$, $z$.
All the Cartesian component of helical pull up satisfies
$$
   |(\hat{u}(\v k)\h{k}{s})_j|
   =    |\hat{u}(\v k)||(\h{k}{s})_j|
   \leq |\hat{u}(\v k)|
$$
so that they are $l^2(\Zz)$-sequence.
Thus the helical pull up is an 
$l^2_{\vss{\chi}}(\Zz)$-sequence by definition.
}

Thus the helical transform is unitary transform acts on 
$l^2_{\vss\chi}(\Zz)$.

\Theorem{convergence of helical vector fields}{
If a scalar sequence $\{u(\v k)\}$ belongs to $l^2(\Zz)$, 
the vector field that is given by 
\eq{\label{eq:helical vector field}
   \v u_s(\v x) := \sum_{\vss k\in\Z3\backslash\{\vss0\}} 
   \hat{u}(\v k) \h{k}{s} \exp(2\pi\i\v k\!\cdot\!\v x)
}
converges in the sense of $L^2_{\vss\chi}(\T3)$-norm for 
each polarity $s=+$, $-$ and  $0$.
}
\Proof{
Applying Riesz-Fisher theorem to each component, they are 
proved to  belong to $L^2(\T3)$.
Thus the vector field  belongs to $L^2_{\vss\chi}(\T3)$ by 
definition.
}

   We call the vector field expressed by 
Eq.(\ref{eq:helical vector field}) {\it s-mode helical 
vector field}.
    This theorem allows us to define the function subspaces 
of $L^2_{\vss\chi}(\T3)$ by
\eq{
   L^2_s(\T3):=\left\{
      \sum_{\vss k\in\Z3\backslash\{\vss0\}} 
      \hspace{-6pt} \hat u(\v k) \h{k}{s}
      \exp\left(2\pi\i\v k\cdot\v x\right) 
      \mbox{{\LARGE  ; }}\left\{\hat u(\v k)\right\}
      \in l^2\left(\Zz\right)
   \right\},
}
and the projection operators $\P_s$ that are linear maps from 
$L^2_{\vss\chi}(\T3)$ to $L^2_s(\T3)$ by
\eq{
   \P_{s}\v u(\v x):= 
      \sum_{\vss k\in\Z3\backslash\{\vss0\}} 
      \Bigl(\hat{\v{u}}(\v k)\cdot\cnjg{\h{k}{s}}\Bigr) 
      \h{k}{s} \exp(2\pi\i\v k\cdot\v x),
}
for $s=\Sigma+$,$\Sigma-$,$D$.
    We call the subspaces {\it helical vector spaces} in 
general, and call each one $\Sigma+$-, $\Sigma-$-, $D$-mode 
space, respectively.

The helical vector spaces $L^2_{\Sigma+}(\T3)$, $L^2_{\Sigma-}
(\T3)$ and $L^2_D(\T3)$ are Hilbert spaces w.r.t. the inner 
product $\bigl<*\bigr>_{\vss\chi}$ and 
orthogonal each other, that is, the inner product of 
arbitrary two vectors $\v u$ $\in$ $L^2_s(\T3)$ and $\v v$ 
$\in$ $L^2_{s'}(\T3)$ is zero if the helical mode indices 
do not coincide $s\neq s'$.
    Therefore the following relation holds;
\eq{
   L^2_{\Sigma+}(\T3) \oplus L^2_{\Sigma-}(\T3) \oplus 
   L^2_D(\T3) \subset L^2_{\vss\chi}(\T3).
}
    The orthogonal complement of the direct sum in 
$L^2_{\vss\chi}(\T3)$  is the space of the harmonic functions 
on $\T3$.
    We will discuss it later.
This orthogonal relations are expressed in terms of the 
projection operators by
\eq{
   \P_s \P_{s'}=\delta(s|s') \P_s,
}
for $s$, $s'=\Sigma+$, $\Sigma-$, $D$.

    The theorem also enable us to define the {\it pull up 
operator} $\P^{\dag}_s$ which is a map from the space of 
scalar functions $L^2(\T3)$ to the space of $s$-mode 
helical vector fields $L^2_s(\T3)$ by 
\eq{
   \P^{\dag}_{s} u(\v x):= 
      \sum_{\vss k\in\Z3\backslash\{\vss0\}} 
      \hat u(\v k) \h{k}{s} \exp(2\pi\i\v k\cdot\v x),
}
and their adjoint operators $\P^{\dag *}$ by
\eq{
   \P^{\dag *}_{s}\v{u}(\v x):= 
      \sum_{\vss k\in\Z3\backslash\{\vss0\}} 
      \left(\hat{\v{u}}(\v k)\!\cdot\!\cnjg{\h{k}{s}}\right) 
      \exp(2\pi\i\v k\!\cdot\!\v x),
}
which satisfy the identity
\eq{
   \left<\P^{\dag}_{s}f,\,\v{g}\right>_{\vss\chi}
      = \left<f,\,\P^{\dag *}_{s}\v{g}\right>,
}
for every $f\in L^2(\T3)$ and $\v{g}\in L^2_{\vss\chi}(\T3)$.
    Each pull up operator is unitary in the sense that the 
identity 
\eq{\label{eq:unitary relation}
   \bigl<f,g\bigr> = \cnjg{\hat{f}(\v 0)}\hat g(\v 0) + 
   \bigl<\P^{\dag}_s f,\P^{\dag}_s g\bigr>_{\vss\chi}
}
holds for every $f$, $g$ $\in L^2(\T3)$ and $s=+$, $-$, $0$.
    Let us define {\it the zero-mean function space} $L^2_0
(\T3)$ by 
\eq{
   L^2_0(\T3):=\left\{ u(\v x) \ ;\ u(\v x) 
   \in L^2(\T3),\ \int u(\v x) \d\v x = 0 \right\}. 
}
    The space is Hilbert with respect to the inner product 
Eq.(\ref{eq:inner product}), and the pull up operators become 
the unitary operators from $L^2_0(\T3)$ in the strict sense.
    This unitary nature of the pull up operators allows us 
to construct an orthonormal complete basis of $s$-mode space 
from a scalar function basis.

\Theorem{construction of the helical basis}{
If $\{f_\lambda\;\lambda\in\Lambda\}$ is an orthonormal 
complete basis of $L^2_0(\T3)$ where $\Lambda$ is 
appropriate set of indices, then the pull up of the basis 
to $s$-mode space $\P^{\dag}_s\{f_\lambda\}$ is an 
orthonormal complete basis of $L^2_s(\T3)$ where 
$s=\Sigma+$, $\Sigma-$, $D$.
}
\Proof{
Orthonormality is obvious because of the unitary relation 
Eq.(\ref{eq:unitary relation}).
Completeness is proved as follows: 
if $\bigl<\P^{\dag}_s f_\lambda,\P^{\dag}_s g\bigr>_{\vss\chi}
=0$ for all $\lambda\in\Lambda$, then $g=0$.
Because the unitary relation $\bigl<\P^{\dag}_s f_\lambda,
\P^{\dag}_s g\bigr>_{\vss\chi}=\bigl<f_\lambda,g\bigr>$ holds 
and $\{f_\lambda\}$ is complete.
Therefore $\P^{\dag}_s g=\v 0$ is concluded.
}

The theorem leads to an important corollary.

\Corollary{construction of the solenoidal basis}{
Under the same conditions as the previous theorem, the union 
of the pull up of the basis to $\Sigma+$- and $\Sigma-$-mode 
spaces, $\P^{\dag}_{\Sigma+}\{f_\lambda\}$ $\cup$
$\P^{\dag}_{\Sigma-}\{f_\lambda\}$, is an orthonormal complete 
basis of the function space of the square integrable solenoidal 
vector fields.
}

Finally, we shall discuss the "residual" of the projection 
operators in order to salvage the modes with wavenumber 
$\v 0$, which are left behind the helical modes.
It is easy to see that the function defined by
\eq{
  \v u_h(\v x) := \v u(\v x) - \Bigl( \P_{\Sigma+} + 
  \P_{\Sigma-} + \P_{D} \Bigr) \v u(\v x)
}
is a uniform vector field on $\T3$ for every $\v u(\v x) 
\in L^2_{\vss\chi}(\T3)$.
Because 
\eq{
   \v u_N(\v x) - \Bigl( \P_{\Sigma+} + \P_{\Sigma-} + 
   \P_{D} \Bigr) \v u_N(\v x) = 
   \hat{u}_x(\v 0)\v e_x + 
   \hat{u}_y(\v 0)\v e_y + 
   \hat{u}_z(\v 0)\v e_z
}
for every elements of the fundamental sequence of 
$\v u(\v x)$.
We will denote the space of uniform vector fields by 
$L^2_H$ and define the projection operator $\P_H$ from 
$L^2_{\vss\chi}$ to $L^2_H$ by
\eq{
   \P_H\v u(\v x):= 
      \hat{u}_x(\v 0)\v e_x + 
      \hat{u}_y(\v 0)\v e_y + 
      \hat{u}_z(\v 0)\v e_z.
}
By definition, $L^2_H(\T3) \cap L^2_s(\T3) = \{\v 0\}$, for 
$s=\Sigma+$, $\Sigma-$, $D$.
Every elements of $L^2_H(\T3)$ is a harmonic function on 
$\T3$, i.e. satisfies $\nabla\cdot\v h=0$, $\nabla\times\v h=
\v 0$.
The degree of freedom of $L^2_H(\T3)$ is three.
It is known that the number coincides with {\it the first 
Betti number} of the manifold $\T3$\cite{massey}.

At this point, the orthogonal decomposition of 
$L^2_{\vss\chi}(\T3)$ is completed.

\Theorem{Hodge-Beltrami decomposition}{
\eq{
   L^2_{\vss\chi}(\T3) = L^2_{\Sigma+}(\T3) \oplus 
   L^2_{\Sigma-}(\T3) \oplus L^2_D(\T3) \oplus L^2_H(\T3).
}
}
This is the synthesis of a special case of Hodge 
decomposition and the Beltrami decomposition.

Let us finish this section by giving a theorem to construct 
the orthonormal complete basis of {\it divergence-free} 
vector fields.
For this purpose, we introduce a triad of uniform vector 
fields \{ $\h{0}{\Sigma+}$, $\h{0}{\Sigma-}$, $\h{0}{D}$ \}.
Choice of the vector fields is arbitrary except for the 
requirement that orthonormal relations $\cnjg{\h{0}{s}}
\cdot\h{0}{s'}=\delta(s|s')$ must be hold.

\Theorem{construction of the divergence-free basis}{
If $\{f_\lambda\}$ is an orthonormal complete basis of 
$L^2(\T3)$, then 
\eq{
  \left\{
    \hat{f}_\lambda(\v 0) \h{0}{\Sigma+} + 
    \P^{\dag}_{\Sigma+} f_\lambda(\v x)
  \right\} \cup \left\{
    \hat{f}_\lambda(\v 0) \h{0}{\Sigma-} + 
    \P^{\dag}_{\Sigma-} f_\lambda(\v x)
  \right\} \cup \left\{
    \h{0}{D}
  \right\}
} 
is an orthonormal complete basis of {\it divergence-free 
vector fields} $L^2_{\Sigma+}(\T3) \oplus L^2_{\Sigma-}(\T3) 
\oplus L^2_H(\T3)$.
}

\section{helical basis}

    Here we gives an anthology of the 
properties which are satisfied for every helical basis.

    First, we describe the procedure to obtain the expansion 
coefficient of the helical basis.
    Consider the helical basis which is obtained by the 
helical pull up of an orthonormal basis of $L^2_0(\T3)$, 
say wavelet basis 
$\{\psi_\lambda;\lambda\in\Lambda\}$ where $\Lambda$ is 
an appropriate set of indices.
    The expansion coefficients of an $L^2_{\vss\chi}$-vector 
field $\v u(\v x)$ with respect to $\P^{\dag}_{s}\bigl\{
\psi_\lambda\}$, the helical pull up of $\{\psi_\lambda\}$ 
are given by the inner products
\ea{
   u_{\lambda,s} & := &
     \bigl<\, \P^{\dag}_s\psi_\lambda(\v x),\,
        \v u(\v x) \,\bigr>_{\vss\chi}
\nonumber\\ & = &
     \bigl<\, \psi_\lambda(\v x),\,
        \P^{\dag *}_s\v u(\v x) \,\bigr>
\nonumber\\ & = &
     \sum_{\vss k\in\Z3\backslash\{\vss0\}} 
     \cnjg{\hat\psi_\lambda(\v k)} \, \Bigl( \hat{\v{u}}(\v k) 
     \cdot \cnjg{\h{k}{s}} \Bigr),
}
for $\lambda\in\Lambda$.
    The last equation leads to the following procedures;
(1) calculate $\F\v u$, the vector-valued Fourier 
transform of $\v u(\v x)$:
(2) calculate the inner product $\hat{\v{u}}\!\cdot\!\cnjg
{\v h_s}$, and we obtain $s$-mode helical Fourier 
coefficients $\{\hat{u}_s\}$:
(3) calculate $\F^{-1}\{\hat{u}_s\}$, the scalar-valued 
inverse Fourier transform of the $s$-mode helical Fourier 
coefficients:
(4) applying the procedure of the scalar-valued function 
expansion, for example, the fast wavelet 
transform algorithm to $\F^{-1}\{u_s\}$.
    Thus we obtain the expansion coefficients 
$\bigl\{ u_{\lambda,s} \bigr\}$.
    It should be remarked here that the procedure given 
above requires no novel and specific algorithm, but only 
such tools as FFT and  FWT, which are popular in the 
signal processing.

    The helical pull up of a real valued scalar function 
$f(\v x) \in L^2(\T3)$ is also a real valued vector field.
    Because the helical vector satisfies 
\eq{
   \h{k}{s}=\cnjg{\h{-k}{s}}
}
for each $\v k\in\Z3$ so that each component of the helical 
pull up of the Fourier coefficient $\hat f(\v k)$ satisfies
\eq{
   \left( \hat f(\v k)\h{k}{s} \right)_j 
      =  \cnjg{\left(\hat{f}(-\v k) \h{-k}{s}\right)_j}
}
for $j=1$, 2, 3 and each $\v k\in\Zz$.
    Thus each element of the helical basis pulled up from a 
real-valued function basis of $L^2(\T3)$ is also real-valued.

    The continuity and differentiability of each element of 
a helical basis depends on those of the original scalar 
function basis.
    These properties of the helical basis, however, may not 
be identical to those of original one.
    Since the multiplication of the polarity vector $\h{k}{s}$ 
lead to the 'twisting' of the values of the scalar function 
to each components.
    In order to evaluate the properties, here we go via 
such a path that goes through the Sobolev's imbedding theorem.

    When the function $f(\v x)$ belongs to the Sobolev space 
$H^{r}(\T3)$, i.e. $f(\v x)$ satisfies
\eq{
   ||f(\v x)|H^r(\T3)||^2=
      \sum_{\vss{k}\in\Z3} (1+|k|^2)^{\frac{r}{2}}|\hat f(\v k)|^2
   <\infty,
}
each component of its helical pull up satisfies 
\eq{
   \bigl|\bigl|\bigl(
      \P^{\dag}_{s}f(\v x)\bigl)_j
      \bigl|H^r(\T3)
   \bigl|\bigl|^2=
      \sum_{\vss{k}\in\Z3} (1+|k|^2)^{\frac{r}{2}}
      |\hat f(\v k)|^2
      |(\h{k}{s})_j|^2
   < ||f(x)|H^r(\T3)||^2
}
for $j=x$, $y$ and $z$.
    This convergence property and the Sobolev's embedding 
theorem guarantees the continuity of the helical pull up of 
$f(\v x)$ when $s>\frac32$.
    For this case, the formula of the helical pull up 
$\P^{\dag}_sf(\v x)$ converges pointwise.
    The differentiability of the helical pull up of $f(\v x)$ 
is guaranteed up to $r$-th order when $f(\v x) \in 
H^{s+\frac32}(\T3)$ where $s \geq r$.

    Next we consider the relation between the $\Sigma+$-mode 
and the $\Sigma-$-mode associated with exchange of parity,
ie reversion of orientation of the coordinate system on $\T3$.
    The parity exchange operator is denoted by $\Gamma$ here.
    When $\Gamma$ acts on a real-valued scalar function, 
say $u(\v x)$, the relation 
\eq{
   \Gamma u(\v x)
   =
   \sum_{\vss k\in\vss\Zz}\hat u(\v k) \exp(2\pi\i\v k\cdot(-\v x))
   =
   \sum_{\vss k\in\vss\Zz}\cnjg{\hat u(\v k)} \exp(2\pi\i\v k\cdot\v x)
}
holds.
    Thus the mirror image of a real-valued scalar function 
$u$ is given by the inverse Fourier transformation of the 
complex conjugate of the Fourier transform of $u$, that is 
\eq{
   \Gamma u(\v x)=\F^{-1} \left( \cnjg{\F u} \right)(\v x).
}
    The mirror image of the helical pull up of a real-valued 
scalar function is given by
\ea{
   \Gamma\P^{\dag}_{\Sigma\pm}u(\v x)
   & = &
   \sum_{\vss k\in\vss\Zz}\hat u(\v k) \h{k}{\Sigma\pm}
   \exp(2\pi\i\v k\cdot(-\v x)),
\\
   &=&
   \sum_{\vss k\in\vss\Zz}\cnjg{\hat u(\v k)} \h{k}{\Sigma\mp}
   \exp(2\pi\i\v k\cdot\v x).
}
This leads to the commutation relation between the parity 
exchange and the helical pull up for real-valued functions:
\eq{
   \Gamma \P^{\dag}_{\Sigma\pm} = 
   \P^{\dag}_{\Sigma\mp} \Gamma.
}

\section{multiresolution approximation of the vector fields}

In this section, the properties of the MRA of $L^2_0(\T3)$, 
which is defined by the restriction of the MRA of $L^2(\T3)$
to $L^2_0(\T3)$, and its helical pull up's are discussed.
The omission of constant functions from the MRA causes the 
absence of such solenoidal basis that is homogeneous, i.e. 
given by the orbit of a function by the action of finite 
group.
It is discussed that this "discrepancy" is consistent with 
the practically natural postulation for approximating the 
constant vector field on the finite number of grid points.

    First, we briefly review the multiresolution approximation 
(MRA) of the spaces of the functions of period 1.
    For details, one should consult \S3.11 of 
Ref.\cite{meyer1992}.
    The MRA of the function spaces on $\T{}$ is obtained 
by the periodification of the wavelet functions.

    Consider the $r$-regular MRA of $L^2(\R{})$, denoted by 
$\bigl\{ V_j(\R{});\, j\in\Z{} \bigr\}$.
    The completion of each subspaces $V_j(\R{})$ with respect 
to the $L^\infty$-norm, $\clos{L^{\infty}(\R{})}{V_j(\R{})}$,
retains the relation $ f(x) \in V_j \thatis f(2x) \in V_{j+1}$, 
which is one of the MRA conditions.
Let us define $V_j(\T{})$ by 
\eq{
   V_j(\T{}):=\left\{ f(x);
     \,f(x)\in\clos{L^{\infty}(\R{})}{V_j(\R{})},
     \,f(x+1)=f(x)
   \right\}.
}
    The nested sequence $\bigl\{V_j(\T{});\,j\in\Z{}\bigr\}$ 
is called {\it the $r$-regular multiresolution approximation 
of $L^2(\T{})$}.

According to the lemma 13 in \S3.11 of Ref.\cite{meyer1992},
the spaces $V_j(\T{})$ have the following properties.
If $ j \leq 0 $, they are identical.
The space $V_0(\T{})$ consists of constant functions.
The dimension of the space $V_j(\T{})$ is $2^j$.

Let $W_j(\T{})$ be the orthogonal complement of $V_{j}(\T{})$ 
in $V_{j+1}(\T{})$, the space $L^2(\T{})$ is represented in 
terms of a direct sum of the subspaces as follows:
\eq{
   L^2(\T{})= V_0(\T{}) \oplus
   W_0(\T{}) \oplus W_1(\T{}) \oplus W_2(\T{}) \oplus ....
}
    In the present study, we will define the multiresolution 
approximation of $L^2(\T3)$ by the tensor product of the 
one-dimensional ones:
\eq{
   V_j(\T3) := 
   V_j(\T{})\otimes V_j(\T{}) \otimes V_j(\T{})
   \ \ \ \mbox{for}\ j=0,1,2,....
}
    The corresponding spaces $W_j(\T3)$ ($j\geq0$) are 
defined by the orthogonal complement of $V_{j}(\T3)$ in 
$V_{j+1}(\T3)$.
According to the construction procedures, $V_0(\T3)$ consists 
of the constant functions.
The dimension of $V_j(\T3)$ and $W_j(\T3)$ are $2^{3j}$
and $7\cdot2^{3j}$, respectively.

    Now we will try to swim away from the shore of the 
established MRA theory in order to find out the hidden 
bank of the MRA theory of the three-dimensional vector 
field.

    In section three, we introduced the zero-mean 
function space $L^2_0(\T3)$ in order to construct the 
solenoidal function space by helical pull up.
    In terms of the MRA of $L^2(\T3)$, one can obtain a 
nested sequence of the zero-mean subspaces of $L^2(\T3)$, 
$\bigl\{\tilde V_j(\T3)\bigr\}$, defined by 
\eq{
   \tilde V_j(\T3)=\bigoplus_{i=0}^{j-1} W_i(\T3) \ \ \ 
   \mbox{for } j=1,2,3,...,
}
or equivalently defined by {\it the orthogonal complement 
of $V_0(\T3)$ in $V_{j}(\T3)$}.
    The obtained sequence retains the following conditions 
of MRA by definition:
\ea{&&
   \tilde V_j(\T3) \subset \tilde V_{j+1}(\T3) \ \ 
   \mbox{ for } ^\forall j\geq1;
\\&&
   f(x) \in \tilde V_j(\T3) \thatis f(2x) \in 
   \tilde V_{j+1}(\T3) \ \ \mbox{ for } ^\forall j\geq1;
\\&&
   \clos{L^2(\T3)}{ \bigoplus_{j=1}^{\infty} \tilde V_j(\T3) } 
   = L^2_0(\T3).
}
    It is easy to see that the dimension of 
$\tilde V_j(\T3)$ is $2^{3j}-1$.
    The last condition is easy to prove.
    Because all the functions in the space are, by 
definition, orthogonal to all the constant functions that 
belong to $V_0(\T3)$, that is, 
$\int_{\T3}f(\v x)\d\v x=0$ for 
$^\forall f(x) \in
\clos{L^2(\T3)}{ \bigoplus_{j=1}^{\infty} \tilde V_j(\T3) }$. 
    Hereafter we will call the nested sequence $\bigl\{
\tilde V_j(\T3)\bigr\}$ {\it the MRA of $L^2_0(\T3)$} or 
{\it the zero-mean MRA of $L^2(\T3)$}.

    In the previous section, we proved that the helical pull 
up operators are unitary in the strict sense when they act 
on $L^2_0(\T3)$, instead of $L^2(\T3)$.
    Thus the nested sequences that are given by the helical 
pull up of the zero-mean MRA of $L^2(\T3)$,
$\bigl\{\P^{\dag}_{s}\tilde V_j(\T3)\bigr\}$, satisfies 
the conditions
\ea{&&
   \P^{\dag}_{s}\tilde V_j(\T3) \subset 
   \P^{\dag}_{s}\tilde V_{j+1}(\T3) \subset 
   L^2_s(\T3)\ \ 
   \mbox{ for } ^\forall j \geq 1;
\\&&
   \v u(\v x) \in \P^{\dag}_{s} \tilde V_j(\T3) \thatis 
   \v u(2\v x) \in \P^{\dag}_{s}\tilde V_{j+1}(\T3) \ \ 
   \mbox{ for } ^\forall j \geq 1;
\\&&
   \clos{L^2_{\vss\chi}(\T3)}
   { \bigoplus_{j=1}^{\infty} \P^{\dag}_{s} 
   \tilde V_j(\T3) } = L^2_s(\T3),
}
for $s=\Sigma+$, $\Sigma-$ and $D$.
    We will call each nested sequence {\it the MRA of 
$L^2_s(\T3)$} or {\it the s-mode helical MRA}.
    Because the pull up operator is unitary, the dimension of 
$\P^{\dag}_{s}\tilde V_j(\T3)$ is $2^{3j}-1$ for $s=\Sigma+$, 
$\Sigma-$ and $D$.

Intentionally we do not discuss the Riesz basis condition, 
which is one of the properties that constitutes the 
definition of multiresolution approximation, and $r$-regular 
condition till now.
On the latter condition we will discuss in the next section.

In order to discuss the former condition, we define the term 
{\it homogeneous basis}.
Consider a Hilbert space on $\T{m}$, say $H$, and the finite 
group of residues $k/N$ modulo 1, $\Gamma_N$, where $m$, $N
\in\v N$ and $k\in\Z{m}$.
We will say $H$ has {\it homogeneous basis} if there is such 
a natural number $N$ and a function $\phi(x)$ whose orbit 
under the action of $\Gamma_N$, 
$\{\phi(x-k/N)\,;\,k\in\Z{m}\}$ 
is orthonormal complete basis of $H$.
For example, each $V_j(\T{})$ has homogeneous basis which is 
given by the orbit of the scaling function $2^{j/2}
\phi(2^jx)$ by $\Gamma_{2^j}$.

The dimension of each subspace of zero-mean MRA of $L^2(\T3)$ 
$\tilde{V}_j(\T3)$ is $2^{3j}-1$.
Because there is no such an integer $p$ that satisfies 
$2^{3j}-1=p^3$, $\tilde{V}_j(\T3)$ never has any homogeneous 
basis.
So is its helical pull up $\P^{\dag}_s\tilde{V}_j(\T3)$
\Proposition{absence of the homogeneous basis}{
    Each subspace of the MRA of $L^2_s(\T3)$,
$\P^{\dag}_{s}\tilde V_j(\T3)$ ($s=\Sigma+$, $\Sigma-$, 
$D$ and $j\in{\bf N}$), cannot have 
any homogeneous basis.
}

Intuitively this is a consequence of a practical requirement 
for the signal processing or the numerical analysis that the 
vector field $\v u(\v x)$ should be regarded as an element of 
$L^2_H(\T3)$ if all of its values on the grid points located 
at $\v x=\v k/2^j$ are identical.
In mathematical words, if there exists a vector-valued 
homogeneous basis $\{\v\phi_{j\vss{l}}(\v x)\}$, where 
$j$ and $\v{l}$ are indices for the resolution and the 
location, respectively, it should be able to approximate 
not only the elements of $L^2_s(\T3)$ but also those of 
$L^2_H(\T3)$.
Thus the requirement is restated as the following 
postulation.
\Postulation{homogeneous approximation}{
If $\v u(\v x)$ is approximated as $\v u_j(\v x)= \sum c_j
\v\phi_{j\vss{l}}(\v x)$, where $c_j$ is a constant for each 
finite resolution class $j$, i.e. the inner products 
$c_{j,\vss{l}}=\bigl<\v\phi_{j\vss{l}},\v u\bigr>_{\vss\chi}$ 
are dependent only on $j$ and independent of $\v l$, 
$\v u$ belongs to $L^2_H(\T3)$.
}
Defining the helical pull up of the scaling function of 
$j$-th resolution $\phi_j(\v x)$ by
\eq{
   \v{\phi}_{j,s}(\v x):=
      \hat\phi(\v 0)\h{0}{s} + \P^{\dag}_s \phi(\v x),
}
where $\h{0}{s}$ is an arbitrary uniform vector field with 
amplitude one, one can obtain the set of function that is 
given by $\{\Gamma_{2^j}\v{\phi}_{j,s}(\v x)\}$.
It approximates any constant vector field in such a way 
that is stated above.

\section{singularity of $\Sigma\pm$-mode helical vectors and 
spatial coherence of solenoidal helical wavelet}

In this section the spatial coherence of helical wavelet is 
discussed.

MRA is called {\it r-regular} when the generating scaling 
function $\psi_0$ and the associated mother wavelets $\psi_\ep$ 
satisfy the conditions
\eq{
   \forall m\in\Z{},\,\exists C_m < \infty\ \mbox{ s.t. }
   \left|\partial^\alpha\psi_\ep(x)\right| \leq
   C_m (1+|x|)^{-m}
}
for every multi-index $\alpha$ satisfying $|\alpha| \leq r$.
Our interest here is whether the {\it regularity} of the 
helical wavelet suffers from the singularities of the 
$\Sigma\pm$-mode helical vectors.
For this purpose, in this section we will discuss spatial 
coherence in terms of the vector valued function on $\R3$ 
defined by the Fourier integral of $L^2(\R3)$ function:
\eq{
   \v\psi_{\lambda s}(\v x)
      =\int_{\R3}
       \bigl(\F\psi_{\lambda}\bigr)(\v k)
       \h{k}{s} \exp(2\pi\i\v k\!\cdot\!\v x)
       \d\v k.
}
We will call the integral transform {\it integral helical 
pull up (to s-mode)} hereafter.
Vector fields on $\T3$, which we have discussed, are given 
by periodification of the function.
We will  assume $\bigl(\F\psi_{\lambda}\bigr)(\v 0)=0$ in 
order to avoid the arbitrariness of the definition at $\v k
=\v 0$.

Our afraid is as follows.
In order to construct solenoidal basis by helical pull 
up from an orthonormal complete scalar function basis 
$\{\psi_{\lambda};\lambda\in\Lambda\}$, the conditions 
\ea{&&
   \v k\cdot\h{k}{\Sigma\pm}=0,
\\&&
   |\h{k}{\Sigma\pm}|=1,
\\&&
   \sum_{\lambda \in \Lambda} 
   \bigl|\,\F\psi_{\lambda}(\v k)\,\bigr|^2 = 1,
}
are required for every $\v k$.
The first condition says that the vector field is tangential to 
spheres $|\v k|=const.$.
It is a well known result of the differential topology that 
two-sphere $S^2$ is not parallelizable manifold so that 
the vector fields on $S^2$ have at least two singular 
points\cite{adams}.
Thus $\h{k}{\Sigma\pm}$ must have singular points for every 
$|\v k|$'s.
The second condition says that the vector fields should behave 
like the function sgn($x$) around the singular points.
The third condition together with the two previous ones 
leads to a conclusion that there must exist such functions 
that have singular points in their supports.
Therefore, even if $\psi_\lambda$ belongs to the Schwartz 
class ${\cal S}(\R3)$, the vector function $\F\psi_\lambda
\v{h}_{\Sigma\pm}$ breaks the continuity and does not belong 
to ${\cal S}$ when it has singular points in its support.
Furthermore, these properties leads to an expectation that 
the inverse Fourier transform of $\F\psi_\lambda\v{h}_{\Sigma
\pm}$ has algebraic tail like the Fourier transform of sgn 
function.

Asymptotic analysis of the Fourier transform of $\Sigma
\pm$-mode helical vector at large $r$ shows the far field 
behaviour as $r^{-2}$ around $z\sim0$ plane (see Appendix A).
The same exponent of algebraic tail is obtained by a simple 
scaling argument and the two-dimensional distribution of the 
tail is required by the convergence of wavelet function 
under the periodification operation (see Appendix B).

Thus it is plausible that, in the orthonormal solenoidal 
basis on $\R3$ which is constructed by the integral helical 
pull up of a scalar basis, there exist such functions that 
have algebraic tail.
Thus we conjecture that there exist no such orthonormal 
solenoidal wavelet basis that all the species of wavelets 
decreases rapidly at large $r$.

In the preceding part of this section we will examine the 
behaviour at large $r$ numerically using Littlewood-Paley 
type MRA and its helical pull up.

According to  the recipe by Yamada and Ohkitani \cite{yamada1991},
 the Fourier image   $\hat\phi(k)$ of the scaling function $\phi(x)$ is 
given by 
\eq{
   \hat\phi(k)=\sqrt{g(k)g(-k)}, 
}
where 
\eq{
   g(k) = \frac{
      h(\frac{2}{3}-k)
   }{
      h(k-\frac{1}{3}) + h(\frac{2}{3}-k)
   },\ \ \
   h(k)=\left\{\arr{ll}{
      \exp(-k^{-2})  & k > 0,
   \\
      0 & k \leq 0.
   }\right.
}
By definition, this scaling function is
 infinitely differentiable and 
following properties are satisfied: 
\eq{
  \left\{\arr{ll}{
    \hat{\phi}(k) = 1 & \ \ \mbox{ for }
         -\frac13\leq k \leq \frac13, \\
    0 < \hat{\phi}(k) < 1 & \ \ \mbox{ for }
         -\frac23 < k < -\frac13, \ 
          \frac13 < k <  \frac23, \\
    \hat{\phi}(k) = 0 & \ \ \mbox{ for }
         k \leq -\frac23,\ \frac23 \leq k, 
  }\right.
}
and 
\eq{
   \sum_{j\in\vss{Z}}\left|\hat\phi(k+j)\right|^2=1.
}
Then, the mother wavelet  $\psi(x)$ is obtained  by  
 the Fourier transform of the image,  
\eq{
   \hat\psi(k)=
   \sqrt{
     \hat\phi\left(\frac{k}{2}\right)^2
     - \hat\phi\left(k\right)^2
   }
   \exp\bigl( - \pi\i k ).
}
Three dimensional wavelet functions are constructed by 
 tensor products of one dimensional ones. 
 There are  seven species  of mother wavelets 
 $\psi_{\epsilon}$ ($\epsilon = 1,2,...,7$) given by 
\eq{
   \psi_{\ep}(x,y,z) = 
   \psi_{\xi}(x)\psi_{\eta}(y)\psi_{\zeta}(z), \ \ (\xi,\  \eta,\  \zeta = 0\ {\rm or}\ 1), 
}
where    $\psi_0(x):=\phi(x)$ is the one dimensional scaling function, 
 $\psi_1(x):=\psi(x)$ the one dimensional mother wavelet,
 and the label $\ep$ is determined as $\ep=\xi+2\eta+4\zeta$. 

It is obvious from the definition given above that Meyer's 
wavelet belongs to Schwarz class.
Because its Fourier transform has compact support, are 
bounded and infinitely differentiable.
Thus the Littlewood-Paley type MRA (Meyer's wavelet) is 
$r$-regular for any positive integer $r$.

According to our definition of helical vectors, $x$ and $y$ 
components of type $\ep=4$ helical Meyer wavelets $\P^{\dag}
_{\Sigma\pm}\psi_0(x)\psi_0(y)\psi_1(z)$ have singular points 
in their support and do not belong to Schwarz class.
Spatial coherence of these two components are not obvious.
It should be remarked that the Fourier transform of these 
components remain bounded and compact supported so that 
Sobolev's imbedding theorem guarantees the analyticity of 
them.
All the components of helical mother wavelets except these 
two retain the $r$-regular conditions.

    In order to evaluate the localization of wavelet functions, 
we introduce {\it coherence spectrum} defined by
\eq{
   I(\v\psi_{j\ep\vss{l}s}(\v x),\v c,r) :=
      \int_0^{2\pi} \!\!\!\! \int_0^\pi 
      \left|\v\psi_{j\ep\vss{l}s}(\v x-\v c)\right|^2 
      r^2 \sin\vartheta \, 
      \d\th\, \d\ph,
}
where the origin of spherical coordinate system $(r,\th,\ph)$ 
is taken at $\v c$.

In Fig.1, the coherence spectra of the type-$\ep$ wavelet 
functions are depicted.
The other parameters that define the wavelet are the 
resolution class $j$, location $\v l$, and helicity $s$.
Those of the tested wavelet are set to $j=6$, $\v l=\v 0$ and 
$s=+$, respectively.
The center of the spectrum $\v c$ is taken at ($2^{-j-1}\xi$,
$2^{-j-1}\eta$,$2^{-j-1}\zeta$) for the type-$\ep$ scalar or 
helical mother wavelet, where scalar Meyer wavelet $\psi
_{6\ep\vss{0}}(\v x)$ takes its maximum.
The function is evaluated numerically on the $256^3$ number 
of grids given by $\v x=(i_x/N,i_y/N,i_z/N)$ where $N=256$, 
$i_x$, $i_y$ and $i_z$ are integers that satisfy $-N/2 < i_x,
i_y,i_z \leq N/2$.
The spectrum is approximated by the sum
\eq{
  \int_{r_j-\Delta}^{r_j+\Delta}
    I(\v\psi_{\lambda}(\v x),\v c,r_j)\d r\sim 
    \sum_{r_j-\Delta\leq|\v x-\v c|\leq r_j+\Delta}
    \frac{|\v\psi_{\lambda}(\v x-\v c)|^2}{N^3},
}
where the radius $r_j$ and the shell thickness $\Delta$ are 
$r_j=j/N$ and $\Delta=1/2N$, respectively.

In Fig.1(a) the spectra of scalar wavelet functions are 
depicted.
All the wavelets show rather exponential behaviour at 
large $r$ with oscillations.
Helical wavelets, except for the type-4 one, show almost 
the same behaviour as the scalar ones (see Fig.1(b)).
The oscillations at large $r$ are slightly modified.
There are two remarkable features in the spectrum of the 
type-4 helical wavelet.
One is that the spectrum decreases more slowly than those of 
the other wavelets.
The functional form of the spectrum seems rather algebraic 
at medium scales about $0.05<r<0.3$.
The other is that the relative amplitude of the oscillations 
is smaller than that of other helical wavelets.
In Fig.1(c), the contribution of each component of type-4 
wavelet to the spectrum is depicted.
It is obvious from the figure that $x$- and $y$-components 
contribute to the features given above.
The $z$-component, on the other hand, shows the same 
rapidly decreasing features as others.

On the whole, the type-4 helical wavelet is less localized 
and less oscillating at large $r$ than the others. 
These numerically investigated features are consistent with 
the results of asymptotic analysis and scaling argument 
given in the appendices.

\section{concluding remarks}
In the present paper, we proved the Hodge-Beltrami decomposition of 
vector fields on $\T3$. 
Then, we have shown the construction procedure of the  
 orthonormal complete basis of the solenoidal fields on $\T3$. 
This procedure has a merit that
 it requires only conventional numerical algorithms 
(ie   fast Fourier transform and fast wavelet transform) 
 and no novel one  in  calculating the wavelet coefficients.  

Based on this procedure, multiresolution approximation (MRA) of vector fields on 
$\T3$ is  constructed. 
 It was  shown that there exists  no scaling function 
 which has close relation to  the assumption that 
 if a vector field is constant in numerical simulation, 
 it can be  regarded as  an approximation of a uniform  field.   
Nonexistence of a single scaling function which generates MRA might 
have relation to the idea of multiwavelet \cite{alpert}.

We have conjectured that, in any divergence-free three-dimensional 
 vector wavelet basis, there must be such kind of wavelet function 
 that behaves algebraically at large $|\v x|$ so that it breaks the 
 $r$-regular condition.
This conjecture has its root in the facts that the helical vector 
 must be tangential to $S^2$ for every point, and that $S^2$ is not 
 a parallelizable manifold. 
We feel that the requirement of exponential decay imposed in the 
 work by Battle and Federbush is inappropriate if it is required 
 for all the functions that constitutes a solenoidal basis 
 \cite{battle1993}. 
In two or four dimensional spaces, existence of the $r$-regular 
solenoidal vector wavelet basis is plausible because $S^1$ 
and $S^3$ are parallelizable.

\newpage

\newpage
\section*{Appendix A: asymptotic analysis of Fourier
transform of helical vectors}

In this appendix Fourier transform of the helical 
vectors is considered.
Since the helical vectors have singularities, there 
exists such type of helical wavelet that does not 
belong to Schwarz class.
In order to estimate the amplitude of helical wavelet 
at large $r$, 
 first we calculate formally the Fourier transform of  
each component, and 
then, asymptotic analysis at large $r$ is carried out. 

In Cartesian coordinates, 
 the basis vectors of the spherical polar coordinate 
system are decomposed as 
\eq{\arr{ccrcrcr}{
  \vr {k} &= 
     &\ds\frac{ \rho \cos\beta }{\sqrt{\rho^2+\zeta^2}}\ \v e_x &+ 
     &\ds\frac{ \rho \sin\beta }{\sqrt{\rho^2+\zeta^2}}\ \v e_y &+ 
     &\ds\frac{      \zeta     }{\sqrt{\rho^2+\zeta^2}}\ \v e_z, 
\vspace{9pt}\\
  \vth{k} &= 
     &\ds\frac{ \zeta \cos\beta }{\sqrt{\rho^2+\zeta^2}}\ \v e_x &+ 
     &\ds\frac{ \zeta \sin\beta }{\sqrt{\rho^2+\zeta^2}}\ \v e_y &- 
     &\ds\frac{       \rho      }{\sqrt{\rho^2+\zeta^2}}\ \v e_z, 
\vspace{9pt}\\
  \vph{k} &= 
     & - \sin\beta\ \v e_x &+ & \cos\beta\ \v e_y. &&
}}
 where $(\rho,\beta,\zeta )$ denotes 
  cylindrical  coordinates. 
%
We consider here the $x$-component of the Fourier transform of 
$\h{k}{\Sigma\pm}$.
By operating the ninety degree rotation around $z$-axis, one can 
obtain the $y$-component of the Fourier transform.
These two components are relevant to the singularity problem 
of wavelet function.

The Fourier transform of the $x$-component of $\vth{k}$
is given by 
\ea{&&
   \int_{\R3} 
      \left(\vth{k}\right)_x 
      \exp\left( 2\pi\i\v k\cdot\v x \right)
   \d\v k
\nonumber \\&=&
   \int_{-\infty}^{\infty} \hspace{-3pt}
   \int_{0}^{2\pi}         \hspace{-6pt}
   \int_{0}^{\infty}       \hspace{-6pt}
      \frac{ \zeta \cos\beta }{\sqrt{\rho^2+\zeta^2}}
      \exp\Bigl( 2\pi\i\bigl(
          \rho r \cos(\beta-\alpha) + \zeta z
      \bigr)\Bigr)
   \rho\, \d\rho\, \d\beta\, \d\zeta
\nonumber \\&=&
   \int_{0}^{\infty}       
   \biggl(
      2 \i \int_{0}^{\infty} \hspace{-3pt}
         \frac{ \zeta \sin(2\pi z \zeta) }
              { \sqrt{\rho^2+\zeta^2} }
      \d\zeta
   \biggr)
   \biggl(
      2 \cos\alpha 
      \int_{0}^{\pi} \hspace{-6pt}
         \cos\beta \exp\bigl( 2\pi\i \rho r \cos\beta \bigr)
      \d\beta
   \biggr)
   \rho\, \d\rho
\nonumber  \\&=&
   - 4 \pi \cos\alpha
   \int_{0}^{\infty}
      \rho^2 K_1( 2\pi z \rho )\, J_1 ( 2 \pi \rho r )\,
   \d\rho, \label{ApdxA01}
}
 where we define the value  at   $-z$ by 
 exchanging the sign of Eq.(\ref{ApdxA01})
 because the integrand is an odd function. 
Similarly, the Fourier image of the $x$-component of 
 $\vph{k}$ is 
\ea{&&
   \int_{\R3} 
      \left(\vph{k}\right)_x 
      \exp\left( 2\pi\i\v k\cdot\v x \right)
   \d\v k
\nonumber \\&=& -
   \int_{-\infty}^{\infty} \hspace{-3pt}
   \int_{0}^{2\pi}         \hspace{-6pt}
   \int_{0}^{\infty}       \hspace{-6pt}
      \sin\beta
      \exp\Bigl( 2\pi\i\bigl(
          \rho r \cos(\beta-\alpha) + \zeta z
      \bigr)\Bigr)
   \rho\, \d\rho\, \d\beta\, \d\zeta
\nonumber \\&=& - 
   \int_{0}^{\infty}       
   \left(
      \int_{-\infty}^{\infty} \hspace{-3pt}
         \exp(2\pi\i\zeta z)
      \d\zeta
   \right)
   \left(
      \int_{0}^{2\pi} \hspace{-6pt}
         \sin(\beta+\alpha)
         \exp\bigl( 2\pi\i \rho r \cos\beta \bigr)
      \d\beta
   \right)
   \rho\, \d\rho
\nonumber \\&=&
   - 2 \pi\, \i\, \delta(z)\, \sin\alpha
   \int_{0}^{\infty}  \hspace{-6pt}
      \rho\, J_1 ( 2 \pi \rho r )\,
   \d\rho. \label{ApdxA02}
}
>From Eq.(\ref{ApdxA01}) and Eq.(\ref{ApdxA02}),  
 one formally obtains the Fourier transform  of  $(\h{k}{\pm})_x$ as 
\ea{&&
   \int_{\R3} 
      \left(\h{k}{\pm}\right)_x 
      \exp\left( 2\pi\i\v k\cdot\v x \right)
   \d\v k 
\nonumber \\&=&
   \sqrt2 \pi\, 
   \int_{0}^{\infty}  \hspace{-6pt}
      \left(
        -2 \rho K_1( 2\pi z \rho )\,\cos\alpha
        \pm \delta(z)\, \sin\alpha
      \right) 
      \rho\, J_1 ( 2 \pi \rho r )\,
   \d\rho.
\label {ApdxA03}
}
Because Dirac's delta function $\delta(z)$ is zero for $|z|>0$ 
and asymptotic behaviour of modified Bessel's function 
$K_1(2\pi\rho z)\sim\exp(-2\pi\rho z)$ for $|z|\gg1$, 
function is well localized in the $z$-direction.
So we focus on the behaviour around $|z|\ll1$.
%
 In the region $2\pi\rho z\ll1$ and  $2\pi\rho r\gg1$, 
 the Bessel functions are approximated as
\eq{
  K_1(2\pi\rho z)\sim\frac{1}{2\pi\rho z},\ \ 
  J_1(2\pi\rho r)\sim\frac{1}{\pi\sqrt{\rho r}}
  \cos\left(2\pi\rho r-\frac{3\pi}{4}\right).
}
Substituting this expression, we obtain the asymptotic 
expression of Eq.(\ref{ApdxA03}) given by 
\ea{&&
   \int_{\R3} 
      \left(\h{k}{\pm}\right)_x 
      \exp\left( 2\pi\i\v k\cdot\v x \right)
   \d\v k 
\nonumber \\&\sim&
   \sqrt{\frac{2}{r}} \, 
   \left(
     - \frac{1}{\pi}{\rm p.v.}\left(\frac{1}{z}\right)\,\cos\alpha
     \pm \delta(z)\, \sin\alpha
   \right) 
   \int_{0}^{\infty}  \hspace{-6pt}
      \sqrt{\rho}\,\cos\left( 
        2 \pi \rho r - \frac{3\pi}{4}\right)\,
   \d\rho.
\label{ApdxA04}
}
Let us investigate the behaviour of 
 the definite integral in Eq.(\ref{ApdxA04}) 
by replacing   the semi-infinite interval $[0,\ \infty)$  by 
 the finite one $[0,\ \rho_{0}]$, i.e. 
\eq{
   I(\rho_0):=
   \int_{0}^{\rho_0}  \hspace{-6pt}
      \sqrt{\rho}\,\cos\left( 
        2 \pi \rho r - \frac{3\pi}{4}\right)\,
   \d\rho.
}
It is easily confirmed that 
\ea{
   I(\rho_0) & = & 
   \int_{0}^{\rho_0}  \hspace{-6pt}
      \sqrt{\rho}\,\cos\left( 
        2 \pi \rho r - \frac{3\pi}{4}\right)\,
   \d\rho 
\nonumber \\&=&
   \frac{1}{2\pi r}\Biggl\{
      \sqrt{\rho_0}\sin\left(
         2\pi\rho_0 r - \frac{3\pi}{4} \right)
   +  \frac{1}{2\sqrt{2r}}\Bigl[
         S\bigl(2\sqrt{\rho_0 r}\bigr)
       + C\bigl(2\sqrt{\rho_0 r}\bigr)
      \Bigr]
   \Biggr\}
\label{ApdxA05}
}
where  $S$ and  $C$ are Fresnel integrals, 
\eq{
  S(x):=\int_0^x\sin\left(\frac{\pi}{2}t^2\right)\d t,\ \ 
  C(x):=\int_0^x\cos\left(\frac{\pi}{2}t^2\right)\d t. 
}
For sufficiently large $x$, they are asymptotically evaluated as follows:
\eq{
   S(x),\,C(x)\sim \frac12
     + O\left(\frac{1}{\sqrt{x}}\right). 
\label{ApdxA06} 
}
Substituting Eq.(\ref{ApdxA06}) into Eq.(\ref{ApdxA05}), one obtains 
\ea{
   I(\rho_0) & = & 
   \frac{1}{2\pi r}\Biggl\{
      \sqrt{\rho_0}\sin\left(
         2\pi\rho_0 r - \frac{3\pi}{4} \right)
   +  \frac{1}{2\sqrt{2r}}\left[
         1 + O\left(\frac{1}{\sqrt{\rho_0 r}}\right)
      \right]
   \Biggr\}
\nonumber \\&=&
   \frac{1}{4\sqrt2\pi r^{\frac32}}
 + \frac{1}{2\pi r}\Biggl[
      \sqrt{\rho_0}\sin\left(
         2\pi\rho_0 r - \frac{3\pi}{4} \right)
      + O\left(\frac{1}{\sqrt{\rho_0 r}}\right)
   \Biggr], 
}
 which leads to the following approximation: 
\ea{&&
   \int_{\R3} 
      \left(\h{k}{\pm}\right)_x 
      \exp\left( 2\pi\i\v k\cdot\v x \right)
   \d\v k
\nonumber \\&\sim&
   \sqrt{\frac{2}{r}} \, 
   \left(
     - \frac{1}{\pi}{\rm p.v.}\left(\frac{1}{z}\right)\,\cos\alpha
     \pm \delta(z)\, \sin\alpha
   \right) 
\nonumber \\&&\times
   \lim_{\rho_0\to\infty}\left\{
   \frac{1}{4\sqrt2\pi r^{\frac32}}
   + \frac{1}{2\pi r}\Biggl[
      \sqrt{\rho_0}\sin\left(
         2\pi\rho_0 r - \frac{3\pi}{4} \right)
      + O\left(\frac{1}{\sqrt{\rho_0 r}}\right)
   \Biggr]
   \right\}. \label{ApdxA07}
}
 The term 
\eq{
   \frac{1}{4\pi r^2}
   \left(
     - \frac{1}{\pi}{\rm p.v.}\left(\frac{1}{z}\right)\,\cos\alpha
     \pm \delta(z)\, \sin\alpha
   \right) 
}
in Eq.(\ref{ApdxA07}) does not depend on $\rho_0$. 
Thus
 we can draw a  conclusion that 
 the Fourier transform of $(\h{k}{\pm})_x$ is a function 
 asymptotically behaves like $(x^2+y^2)^{-1}$ 
 when $z \sim 0$ and  $|\v x| \gg 1$.
\newpage
\section*{Appendix B: some estimations a priori}

In Appendix A we showed the corroboration of the existence 
of algebraic tail whose exponent is $-2$.
The tail is shown to distribute rather two-dimensionally 
around $z\sim0$.
In this appendix we will show that the same conclusion is 
obtained by a simple scaling argument and by the requirement 
of convergence of wavelet function under the periodification 
operation.

Assuming that singularities of the helical vectors cause 
{\it algebraic tail} of helical wavelet at sufficiently 
large $r=|\v x|$, exponent of the tail is determined by a 
simple scaling argument.
When wavelet function in $W_j$ is assumed to be divided 
into two parts, rapidly decreasing part and algebraic tail, 
as
\eq{
  \psi_{\lambda}(\v x)=f(\v x) + A r^{-\alpha},
}
for sufficiently large $r$, the wavelet in $W_{j+1}$ must 
be given by 
\eq{
  \sqrt{2^3}\psi_{\lambda}(2\v x)
    = \sqrt{2^3} f(2\v x) + 2^{\frac32-\alpha} A r^{-\alpha}.
}
The domain of Fourier integral is enlarged by two in each 
direction so that the contribution of the singularities of 
helical vectors to class $W_{j+1}$ wavelet become twice of 
those in class $W_j$.
(Remember that functional form of $\v h
_{\pm}$ around the singular points does not depend on $|\v k|$ 
and the points are aligned with a one-dimensional manifold).
The amplitude of the Fourier transform of $\psi(2\v x)$ is 
one-eighth of that of $\psi(\v x)$.
Thus the factor of algebraic term should be $2^{-\frac12}$, 
one-fourth of the factor of $f$, and the exponent $\alpha=2$.

Scaling argument leads to a conclusion that the exponent of 
algebraic tail is $-2$.
Is such behaviour of $r^{^2}$ isotropically distributed, i.e. 
found in all the directions of three-dimensional space?
The answer is no.
They distributes two-dimensionally because of the following 
two reasons.
One is that the singularity distributes one-dimensionally so 
that the Fourier transform in the directions perpendicular 
to the singularity suffers from it.
The other is that helical wavelet on $\R3$, say $\v\psi_R
(\v x)$, should not diverge under the periodification 
operation to obtain the wavelet on $\T3$.
This requirement leads to an estimation of the contribution 
of algebraic tails from far field to the periodification 
sum:
\eq{
  \sum_{|\vss{l}|>r_0}\v\psi_R(\v x+\v l) 
  \propto \int_{r_0}^{\infty}\frac{1}{r^2}\,
  r^{D-1}\,\d r < \infty  \then D < 2,
}
where $r_0$ is a sufficiently large number and $D$ is the 
(fractal) dimension of region where the algebraic tails 
spread over.
Though the estimation is quite rough, this result claims 
that the tail must not spread over to all the directions of 
three-dimensional space, but should be confined in the 
region which is at most two-dimensional.

\newpage
\section*{Figure captions}

Fig.1 Coherence spectra of wavelet functions; (a) scalar 
Meyer wavelet, (b) helical Meyer wavelet.
Type of wavelet is distinguished by the line; solid line: 
$\ep=1$, 2, broken line; $\ep=3$, 5 and 6, dashed line: $\ep
=4$, dash-dotted line: $\ep=7$.
(c) Spectra of each component of type-4 helical wavelet.
Solid line: $x$-component, dash-dotted line: $y$-component, 
broken line: $z$-component. Solid and dash-dotted lines are 
indistinguishable.
\end{document}